\documentclass{article}

\usepackage{arxiv}

\usepackage[utf8]{inputenc} 
\usepackage[T1]{fontenc}    
\usepackage{hyperref}       
\usepackage{url}            
\usepackage{booktabs}       
\usepackage{amsfonts}       
\usepackage{nicefrac}       
\usepackage{microtype}      
\usepackage{subfig}
\usepackage{multirow}
\usepackage{graphicx}

\title{AI Assisted Annotator using Reinforcement Learning}

\author{%
	V. Ratna Saripalli \\
	GE Healthcare\\
	\texttt{Ratna.Saripalli@ge.com} \\
	\And
	Gopal Avinash\\
	GE Healthcare\\
	\texttt{Gopal.Avinash@med.ge.com} \\
	\And
	Dibyajyoti Pati\\
	GE Healthcare\\
	\texttt{Dibyajyoti.Pati@ge.com} \\
	\And
	Michael Potter\\
	GE Healthcare\\
	\texttt{michael.potter@ge.com} \\
	\And
	Charles W. Anderson \\
	Colorado State University \\
	\texttt{Chuck.Anderson@colostate.edu} \\
}

\date{}

\begin{document}
\maketitle

\begin{abstract}
Machine Learning in the healthcare domain is often hindered by data which is both noisy and lacking reliable ground truth labeling. Moreover, the cost of cleaning and annotating this data is significant since, unlike other data domains, medical data annotation requires the work of skilled medical professionals. In this work, we report on the use of reinforcement learning to mimic the decision-making process of annotators for medical events\textemdash allowing automation of annotation and labeling. Our reinforcement agent learns to annotate health monitor alarm data based on annotations done by an expert. This method shows promising results on such alarm data sets.

\end{abstract}


\section{Introduction}
\label{sec:intro}

Healthcare costs have increased significantly worldwide over the past decades, and especially within the United States where they now account for nearly 18\% of the GDP \cite{healthcosts}. As such, it is imperative to find ways to lower systemic costs while not compromising quality of care. In particular, Operating Room costs represent a significant fraction of a hospital's expenditures. Many sequential decision-making steps are involved in the day-to-day function of an OR. Common examples include deciding when to transfer patients to post-anesthesia care units, scheduling staff for care units, determining surgery end, estimating emergence phase, estimating time to extubate, and setting critical event alarms. These problems present attractive targets for novel machine learning applications which, leveraging an ever-growing corpus of medical data, may be able to significantly streamline hospital logistic operations. 

Data is the fundamental currency for solving many healthcare problems using computational methods. While volumes of medical data are increasingly being made available, these datasets bring their own unique challenges. Medical data is plagued with concerns ranging from data privacy and ground truth availability, to domain issues such as sparsity and heterogeneity, to quality issues such as missing data and noise \cite{zhou2017machine, Xiao2018OpportunitiesAC, ghassemi2018opportunities, esteva2019guide}. Moreover, obtaining trustworthy annotations of medical monitoring data is expensive and time consuming due to domain expertise requirements \cite{schwab2018not}. On the other hand, medical devices such as anesthesia machines, ventilators, and monitoring systems can be a rich source of data as they assist in processing, identifying, and generating alerts for patients. Many such alerts in turn serve as the basis for optimal OR decision making. 

One well-known problem with alerts from conventional medical monitoring systems is alarm fatigue, which can be attributed to the threshold-based alarm classification approaches currently used in such systems \cite{wang2015machine,schwab2018not}. Patients and their families become anxious when alarms indicate a change in the patients’ health and a need for medical attention. An overabundance of false alarms also tends to drive operators to either silence or ignore the alarms\textemdash reducing the utility of the monitoring device. To build smarter alarm systems leveraging modern deep learning algorithms, we must address the problem of false alarms. This requires a significant volume of correctly annotated data. Unfortunately, as noted before, this is both expensive and time consuming \cite{schwab2018not}.

Reinforcement Learning (RL) is a computational approach to learning from interactions that is goal focused and has gained a lot of attention in the last five years \cite{sutton1998introduction}. RL algorithms center around an agent that senses, observes, and interacts with an environment. The environment, in turn, either rewards or penalizes the agent to attain a specific goal. RL is especially useful in the automation of tasks which require human goal oriented action and sequential decision making. These algorithms have seen great success even in complicated tasks such as playing games at human and super-human levels of performance \cite{volodymyr2015human}. We report on a novel application of RL to the domain of medical alarm annotation. In particular, we compare several RL algorithms and feedback mechanisms in an effort to mimic a human expert at the task. 

\section{Related Work}
\label{sec:related}

Electronic Health Record and Electronic Medical Record health systems have matured over the past decades, and yet the data generated from these systems are yet to be tapped to their full potential. Many traditional approaches have been used for detecting false alarms that depend on feature engineering and the availability of ground truth (labeled) data \cite{wang2015machine,sayadi2011life,salas2014false,behar2013ecg,clifford2015physionet,plesinger2016taming}. For example, Wang et al. \cite{wang2015machine} used a three-step approach of feature extraction, selection, and classification based on arterial blood pressure (ABP) and electrocardiogram (ECG) signals for false alarm detection. The authors found that direct raw signals yield poor results due to noise and unstable voltages. To overcome this they developed a customized feature set using statistical methods, which was then passed to a support vector machine (SVM) to classify the alarms. While their work achieved high classification performance, the results are limited to ECG arrhythmia alarms and required significant feature engineering. 

Although such traditional approaches to false alarm detection give reasonable model performance, they suffer from several significant limitations. One chief concern is that they typically focus on a single alarm/signal type (e.g., arrhythmia) and therefore cannot scale and generalize for various alarm types. Another important concern is their poor performance when only a small set of ground truth data is available. Distant supervision\textemdash mapping of entity relations from a known knowledgebase to a dataset that has unlabeled data \cite{mintz2009distant}\textemdash can be used to alleviate the limited ground truth problem, but such methods still do not address the generalization problem \cite{schwab2018not}. In their work, Schwab et al. \cite{schwab2018not} use a multitask network architecture to perform auxiliary tasks and detect false alarms via distant supervision. While this proved to be beneficial when the available labeled dataset was small (under 100 samples), their work was limited to only detecting false alarms due to technical errors or artifacts. It also required the use of auxiliary tasks, which is not always natural or desirable for a given dataset. 

Moving beyond classical methods, RL algorithms have begun to show success in medical domain problems such as defining ventilation weaning protocols and customizing drug administration strategies \cite{prasad2017reinforcement,escandell2014optimization,nemati2016optimal,padmanabhan2015closed}. In 2015, the field of deep RL was born when researchers combined RL techniques with deep neural networks to create the Deep Q-Network (DQN). The DQN was able to solve complex state-space problems in the form of Atari 2600 games, at the level of a professional human player, without any prior domain knowledge \cite{volodymyr2015human}. A more recent RL algorithm, the Advantage Actor-Critic (A2C), learns to approximate both its policy and value functions and then uses the latter to update the former \cite{sutton1998introduction,mnih2016asynchronous}. We see the benefits of A2C's learned value function as compared to the DQN approach in our results (Section \ref{sec:results}).

\section{Methods}
\label{sec:methods}

\begin{figure}[ht]
	\vskip 0.2in
	\begin{center}
		\centerline{\includegraphics[width=\columnwidth]{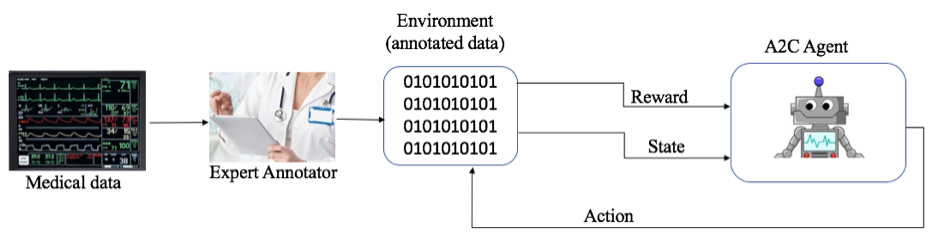}}
		\caption{Overview of a data-driven RL annotation framework for medical events}
		\label{fig:overview}
	\end{center}
	\vskip -0.2in
\end{figure}

We trained DQN and A2C agents to annotate medical signal data based on whether or not it represents an alarm state. The high level data flow is diagrammed in Figure \ref{fig:overview}.

Our proposed RL approach can learn from the decision making of the domain expert without any assumptions regarding signal type nor pre-encoding any domain expertise. Once the RL agent reaches reasonable performance, we can replace the human expert with the RL agent to annotate the data. Depending on accuracy requirements, a production system might leave a human in the loop to validate the annotations output by the RL agent, but their workflow could be accelerated by access to the agents predictions. 

\subsection{Datasets}
\label{sec:datasets}

We used the multi-phasic Push Electronic Relay for Smart Alarms for End User Situational Awareness (PERSEUS) program’s data hosted by Brown University's digital archive \cite{perseus}. This data was generated from an adult Emergency Department (ED) for a regional referral medical facility and level I trauma center. It contains data from patient monitoring devices within a 15-bed urgent care area in the ED. The PERSEUS dataset contains 12 months of .json formatted data. The data are split across files by monitoring device and by 24 hour increments. The following signals are recorded in each file:

\begin{itemize}
	\item Electrocardiogram waveform (single lead EKG , Lead II) at 250Hz
	\item Pulse oximetry waveform (PPG) at 125Hz
	\item Vital signs: heart rate (HR), respiratory rate (RR), systolic blood pressure (SBP), diastolic blood pressure (DBP), mean arterial blood pressure (MAP) and peripheral capillary oxygen saturation (SPO2)
	\item Alarm messages (institution-specified alarms)
	
\end{itemize}

Kobayashi et al., as part of the PERSEUS program, developed several data subsets with annotation for experimental (non-clinical) research known as Adjudicated / Annotated Telemetry signals for Medically Important and Clinically Significant events (ATOMICS), which are used in this research \cite{kobayashi2018development}. From the 12 months of PERSEUS data, three non-consecutive weeks of red alarms data are annotated by Kobayashi et al. for clinical significance and severity to form the ATOMICS dataset. This dataset is then further divided by weeks into ATOMICS-1, ATOMICS-2, and ATOMICS-3. We use ATOMICS-1 as our training data and ATOMICS-2 as our testing data.

\subsection{Preprocessing data}
\label{sec:preprocessing}

\begin{figure}[ht]
	\vskip 0.2in
	\begin{center}
		\centerline{\includegraphics[width=\columnwidth]{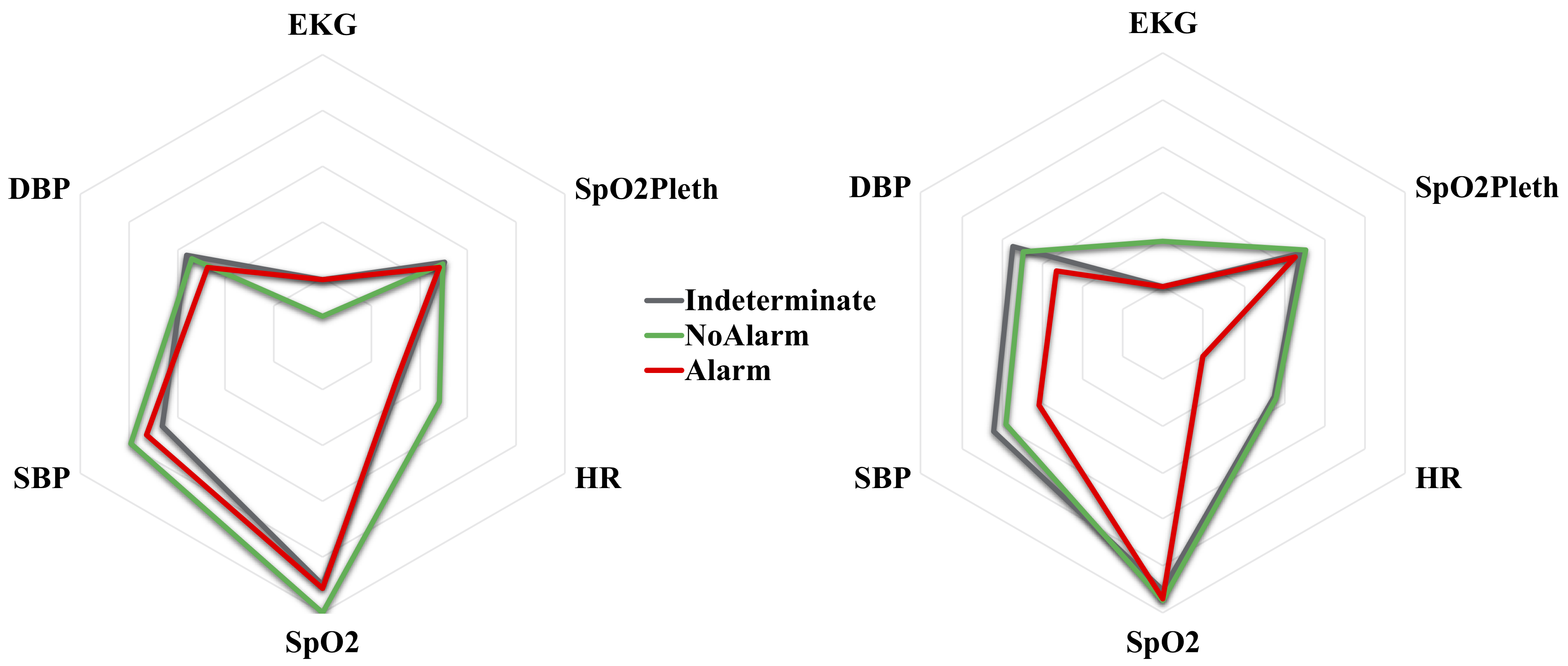}}
		\caption{Two examples of True, False, and Indeterminate alarms}
		\label{fig:alarms}
	\end{center}
	\vskip -0.2in
\end{figure}

We first convert all alarms and annotations to one-hot encodings for processing. The annotations are divided into two categories of actions (alarms/non-alarms) to simplify the problem space. The clinically significant and severe alarms (emergent, urgent) are categorized as alarms, while the indeterminate and non-urgent events are categorized as non-alarms. The raw data from three files\textemdash vitals, alarms, and annotations\textemdash are then merged together into a single file based on timestamps. This results in dataset 1 (DS1) as summarized in Table \ref{tbl:summary}. We then create a second dataset (DS2) from DS1, where `indeterminate' events are removed. These are events where the professional annotator was unable to make a definite assessment of alarm status. Two examples of such events can be seen in Figure \ref{fig:alarms}, where the indeterminate event can be seen to share some attributes of a true alarm and some attributes of a non-alarm. DS2 is also down-sampled from milliseconds to seconds using mean value imputation. Its final data distribution is summarized in Table \ref{tbl:summary}.

The data is highly imbalanced and sparse for critical clinically significant alarm events. Each critical alarm is surrounded by 600 non-alarm events before and after. This results in an imbalance ratio of 1:1200. To rectify this imbalance we used the following downsampling techniques: n-0, n-1, n-3, n-5, n-10, and mixed. In n-0 downsampling, we retain only alarm data. In n-1, we retain 1 non-alarm before and after an alarm. Similarly, in n-3, n-5, and n-10, we retain 3, 5, and 10 surrounding non-alarms respectively. Mixed is a random sampling of these strategies combined. 

\subsection{Problem formulation}
\label{sec:problem}

A Markov decision process for our alarm annotation RL problem is defined as follows: 

\begin{itemize}
	\item $s_t $ is a vector of the six physiological variables described in Section \ref{sec:datasets} at a given time $t$. $S$ is the finite state space comprising the set of $s_t$, with transitions $s_t \to s_{t+1}$.
	
	\item An action space $A$ where an agent takes an action $a_t \in A$ at each time step and the state is changed to $s_{t+1}$. We are using historical data to mimic the environment. In future work, for clinically realistic use-cases, a medical intervention would be required when an alarm goes off. Only after the patient returns to normalcy would the medical staff reset (perform a state transition on) the alarm. In our case $A=\{1,0\}$ corresponding to alarm and non-alarm.
	
	\item For experiments using DS1, at each time step $t$ a scalar reward value of 1 was given for correctly identifying a non-alarm, 10 for an alarm, and 0 for an incorrect choice.
	
	\item For experiments using DS2, at each time step $t$ a reward was computed based on a function inspired by \cite{prasad2017reinforcement}. Our form of the reward function\textemdash a combination of sigmoid, piecewise-linear, and threshold functions based on vitals\textemdash is given in Equation \ref{eq:reward}. 
	
	\begin{equation}
	5 \sum_{v \in V} \left[\frac{1}{1+e^{-50(x_1)}} - \frac{1}{1+e^{-50(x_2)}} + \frac{1}{2} \right]
	\label{eq:reward}
	\end{equation}	
	
	In Equation \ref{eq:reward}, $V$ is a set of three out of the six physiological variables: $V=\{HR, SBP, DBP\}$. $x_1 = (v_t-v_{min})$ and $x_2 = (v_t-v_{max})$. $v_{min}$ and $v_{max}$ are minimum and maximum thresholds for each variable, given in Table \ref{tbl:thresholds}.
	
\end{itemize}

\begin{table}[t]
	\caption{Threshold values for physiological signals}
	\label{tbl:thresholds}
	\vskip 0.15in
	\begin{center}
		\begin{small}
			\begin{sc}
				\begin{tabular}{lccr}
					\toprule
					Signal & $v_{min}$ & $v_{max}$ & Units \\
					\midrule
					HR & 60& 120& BPM \\
					SBP & 90& 200& mmHg\\
					DBP & 60& 140& mmHg\\
					\bottomrule
				\end{tabular}
			\end{sc}
		\end{small}
	\end{center}
	\vskip -0.1in
\end{table}

The goal of the RL agent is to maximize its expected reward by using known examples to learn an optimal policy.

\subsection{Learning an optimal policy}
\label{sec:learning}

Learning the best mapping (Q-function) between actions and states is the essence of reinforcement learning. We compare two different models for learning this mapping: a DQN with experience replay (value-based learning \cite{volodymyr2015human}), and an A2C (policy-based learning \cite{sutton1998introduction,mnih2016asynchronous}).

The DQN network takes in $s_t$ as described in Section \ref{sec:problem}, and it outputs a Q-value for each action $a \in A$ pursuant to Equation \ref{eq:dqn}. The model parameters ($\theta$) are updated after every 10 steps of training within each epoch, with a batch size of 8, learning rate of 0.001, and an Adam optimizer. The discount factor ($\gamma$) is set to 0.9.

\begin{equation}
\hat{Q}_k(s_{t},a_{t})\leftarrow r_{t+1} +\gamma \max_{a \in A} \hat{Q}_{k-1}(s_{t+1},a, \theta)
\label{eq:dqn}
\end{equation}

The optimal policy $\pi^*$ for the DQN method after k iterations is given in Equation \ref{eq:dqn_opt}.

\begin{equation}
\label{eq:dqn_opt}
\pi^*(s)= {argmax}_{a \in A} \hat{Q}_k(s,a)
\end{equation}

The A2C algorithm is comprised of two components: a policy function, and a value function. To update these functions, A2C maintains two neural networks. The critic network predicts the value $V_w(s_t)$ at time $t$ from the state parameters $s_{t}$, and updates the value function parameters $w$, which in turn are used for computing `Advantage' as described in Equation \ref{eq:a2c}.

\begin{equation}
Advantage(s_t,a_t)\leftarrow r_{t+1} +\gamma V_w({s_{t+1}}) - V_w({s_t})
\label{eq:a2c}
\end{equation}

The value function for a given policy is described in Equation \ref{eq:value}.

\begin{equation}
V(s_{t+1})\leftarrow E [\sum\limits_{t=1}^{T}\gamma^{t}r_{t}]
\label{eq:value}
\end{equation}

In Equations \ref{eq:a2c} and \ref{eq:value}, $r_{t}$ is the reward after time $t$. The actor updates the policy per the critic in order to maximize `Advantage' at a given time $t$ from the input state parameters $s_t$.

Both the actor and critic networks are updated every 10 time steps during an epoch, with a batch size of 8. The networks use an Adam optimizer with a learning rate of 0.001 for the actor and 0.005 for the critic. The discount factor ($\gamma$) is set to 0.9. 

Both the DQN and A2C models choose actions based on a decaying $\epsilon$-greedy rule with an initial value of 1 annealed to 0.01 using a decay factor of 0.99975. 

\subsection{Experiment design}
\label{sec:experiment}

All experiments were run on a MacBook Air with an Intel Core i5 1.8 GHz processor and 8 GB of RAM. Training was performed using ATOMICS-1 data, and all agent evaluations were conducted using ATOMICS-2 data. Table \ref{tbl:summary} summarizes the characteristics of the datasets. Standard binary classification metrics such as sensitivity, specificity, and the Mathews Correlation Coefficient (MCC) \cite{mcc} were employed to compare model performance. Code was written using the Keras sequential model API and Python 3.6. 

\begin{table}[t]
	\caption{Summary of ATOMICS datasets used for training and testing}
	\label{tbl:summary}
	\vskip 0.15in
	\begin{center}
		\begin{small}
			\begin{sc}
				\begin{tabular}{lcccr}
					\toprule
					DS1 & True & Non-Alarms & Total \\
					\midrule
					ATOMICS-1-Train & 437& 406& 843 \\
					ATOMICS-2-Test & 756& 468& 1224\\
					\toprule
					DS2 & True & False & Total \\
					\midrule
					ATOMICS-1-Train & 434& 208& 642 \\
					ATOMICS-2-Test & 750& 280& 1030\\
					\bottomrule
				\end{tabular}
			\end{sc}
		\end{small}
	\end{center}
	\vskip -0.1in
\end{table}

\subsection{Benchmarking against non-RL algorithms}
\label{sec:benchmarking_algos}

We also benchmark our RL algorithms against three non-RL based approaches: a Multi-Layer Perceptron (MLP), an SVM, and a recurrent neural network using Long Short-Term Memory (LSTM). The MLP and SVM both take $s_t$ as input and return a classification value in $\{1,0\}$ for alarms and non-alarms. They are benchmarked by training with n-0, n-3, and n-10 downsampling on DS1. The LSTM is fed a sequence of $s_t$ values surrounding the target event, and then returns a classification value. If the event occurs at time $t$, then an LSTM with window size $w$ sees state data in the range $s_{t-w/2}\to s_{t+w/2}$.

The MLP consists of two hidden layers with twenty and four nodes respectively, each using a hyperbolic tangent activation function leading into a softmax output layer. It was trained until convergence for a fair comparison with the RL agent. For n-3 downsampling the class\_weight was set to 20:1, and for n-10 downsampling it was set to 40:1. Without adjusting these weights the classifier collapses to always predicting the majority class.  

We used scikit-learn’s sklearn.svm package, and support vector classifier (SVC) for our SVM experiments. After experimenting with poly, rbf, and linear kernels, we selected the linear kernel as this gave the maximum AUC score. We also set the class\_weight parameter to `balanced'. Without adjusting these weights the classifier collapses to always predicting the majority class. All other parameters were left as their package default values.

Our recurrent neural network consists of four stacked layers with two initial LSTM layers followed by two dense layers. We used a batch-size of 16, the RMSProp optimizer, and a learning rate of 0.001.

\section{Results and Discussion}
\label{sec:results}

The purpose of an AI-assisted annotator is to mimic an expert annotators behavior. Our results are discussed in detail in the following sections. A2C agents have previously been found to learn sparse events more effectively than DQNs, a finding which is borne out here as well. The following sections detail the experimental results from this study. 

\subsection{Comparing the optimizers Adam and RMSProp}
\label{sec:optimizers}

\begin{figure}[ht]
	\vskip 0.2in
	\begin{center}
		\centerline{\includegraphics[width=\columnwidth]{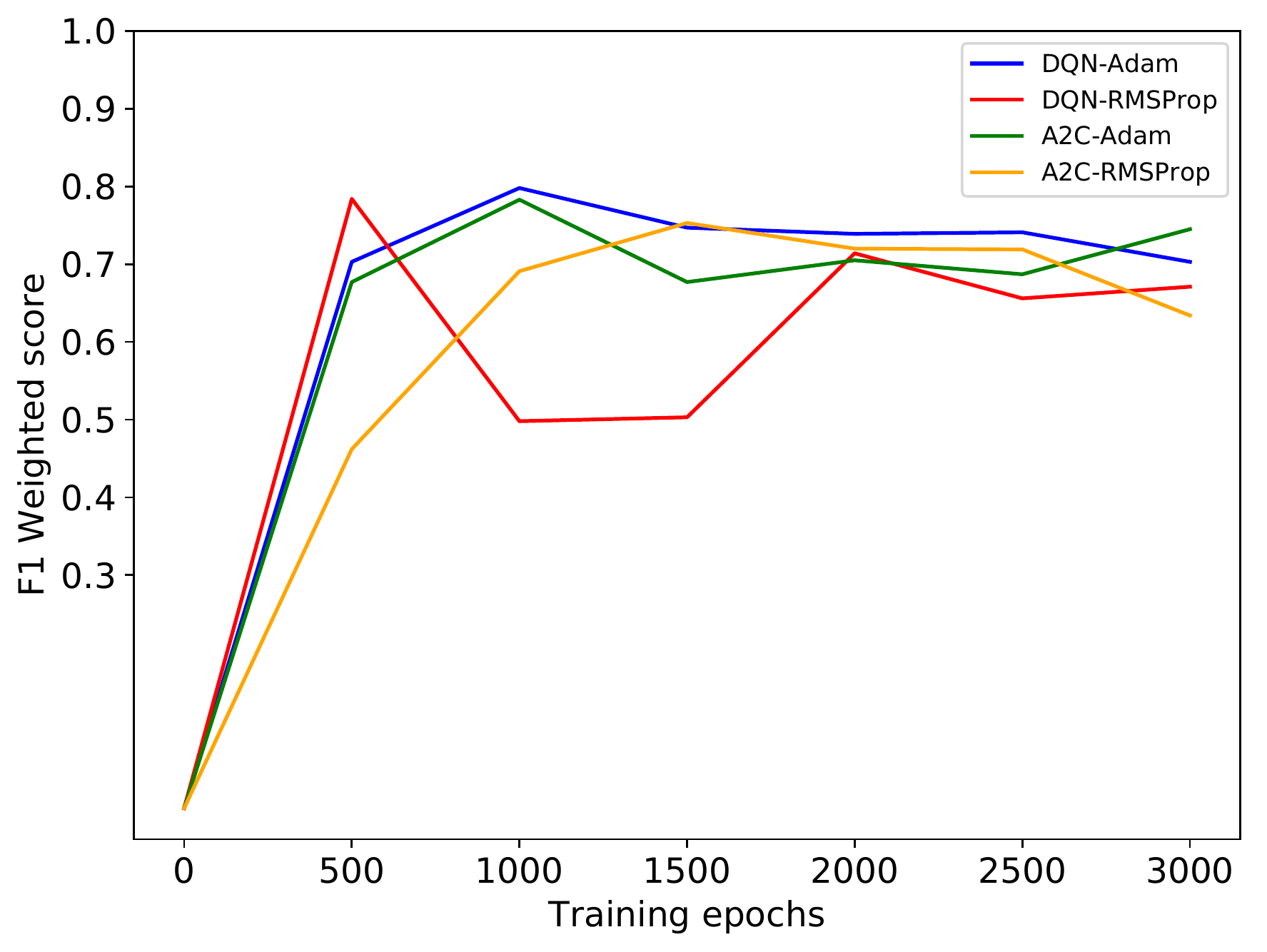}}
		\caption{Comparison of Adam and RMSProp optimizers for the DQN and A2C networks.}
		\label{fig:opt}
	\end{center}
	\vskip -0.2in
\end{figure}

The performance of various agents trained at different epochs was evaluated and compared for DQN and A2C using F1 scores weighted by class prevalence, as seen in Figure \ref{fig:opt}. We used n-mixed downsampling on DS1.

As can be seen from Figure \ref{fig:opt}, the Adam optimizer was found to be more stable than RMSProp. RMSProp performance fluctuated significantly for DQN and resulted in a worse final performance for both DQN and A2C. We therefore continued with Adam for the rest of the experiments. 

\subsection{Comparing the results of A2C and DQN agents}
\label{sec:agents}

In medical alarm annotation, sensitivity is a very important metric since we don't want to miss any true alarms. Specificity is the `value-added' metric which indicates how effectively we can identify false alarms. The performances of our A2C and DQN algorithms are compared with respect to these metrics in Table \ref{tbl:compareagents}, along with AUC and MCC scores. Each agent (A2C and DQN) were run for 5,000 epochs of training, being evaluated against the test dataset every 100 epochs. The top 3 (highest AUC) evaluation results during this training process are reported in Table \ref{tbl:compareagents}. We find that A2C consistently achieves higher sensitivity scores than DQN, at best identifying 89\% of true alarms compared with 76\% in the case of DQN. On the other hand, that A2C network would catch only 42\% of false alarms compared to the DQN's 64\%. While catching more false alarms would be nice, a realistic system shouldn't miss any true alarms, so A2C is preferable. 

\begin{table}[t]
	\caption{Comparing the results of A2C and DQN agents with n-mixed downsampling on DS1. The top three (AUC) results for each algorithm are given, with best results highlighted in bold.}
	\label{tbl:compareagents}
	\vskip 0.15in
	\begin{center}
		\begin{small}
			\begin{sc}
				\begin{tabular}{lccccr}
					\toprule
					Agent & AUC & MCC & Sens. &Spec. \\
					\midrule
					A2C-1  & 0.681 &0.367&0.774& 0.589 \\
					A2C-2  & 0.671 &0.351& 0.791& 0.551 \\
					A2C-3  & 0.662 &0.376& \textbf{0.896}& 0.427 \\
					&&&&&\\
					DQN-1 & \textbf{0.704} &\textbf{0.406}& 0.760& 0.647\\
					DQN-2 & 0.662 &0.335& 0.459& 0.865\\
					DQN-3 & 0.627 &0.290& 0.335& \textbf{0.918}\\
					\bottomrule
				\end{tabular}
			\end{sc}
		\end{small}
	\end{center}
	\vskip -0.1in
\end{table}

\subsection{A2C Agents training curves and reward signals}
\label{sec:curves}

\begin{figure}[ht]
	\vskip 0.2in
	\centerline{\includegraphics[width=\columnwidth]{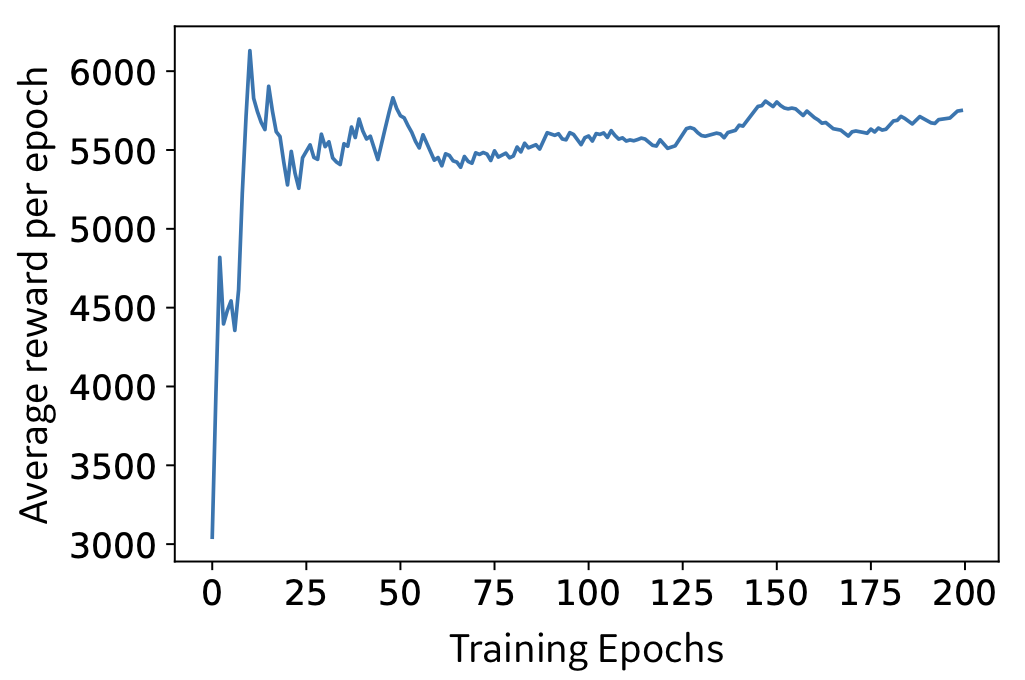}}
	\caption{An A2C training curve tracking the agent's average reward per epoch using n-mixed downsampling on DS1.} 
	\label{fig:a2c}
	\vskip -0.2in	
\end{figure}

The training curve for an A2C agent is provided in Figure \ref{fig:a2c}. This captures the models average accumulated reward each epoch when training with n-mixed downsampling on DS1. As can be seen from the figure, the network steadily learns a better action policy. 

\subsection{Benchmarking RL and ML results}
\label{sec:benchmarking_results}

We benchmarked our top A2C agent results against MLP and SVM binary classification results with n-0, n-3, and n-10 downsampling on DS1 (see Table \ref{tbl:summary}). The results of this benchmarking are presented in Table \ref{tbl:bm1}. In these tests the the SVM mostly outperforms A2C and MLP. It's worth noting, however, that if the MLP and SVM are not adjusted for class imbalances they both collapse to majority-only classifiers. A2C has no explicit class re-balancing, so we also run n-mixed downsampling as a pseudo imbalance rectifier. The result is that the A2C model achieves the highest recorded sensitivity while maintaining a 45\% specificity.

\begin{table}[t]
	\caption{Comparing the performance results of RL and ML methods on DS1}
	\label{tbl:bm1}
	\vskip 0.13in
	\begin{center}
		\begin{small}
			\begin{sc}
				\begin{tabular}{lccccc}
					\toprule
					Range&Algo.&AUC &MCC&Sens. &Spec. \\
					\midrule
					&A2C& 0.641 &0.306& \textbf{0.828}& 0.455 \\
					n-0 &MLP& 0.674 &0.351& 0.581& 0.767 \\
					& SVM& \textbf{0.708} &\textbf{0.454}& 0.528& \textbf{0.888} \\
					&&&&  \\
					&A2C& 0.661 &0.319& 0.532& 0.790 \\
					n-3 &MLP& 0.614 &0.324& 0.278& \textbf{0.951} \\
					& SVM& \textbf{0.723}&\textbf{0.469} &\textbf{0.585} & 0.861 \\
					&&&&  \\
					&A2C& 0.553 &0.171& 0.142& \textbf{0.963} \\
					n-10 &MLP& 0.691 &0.429& 0.482& 0.898 \\
					& SVM & \textbf{0.723}&\textbf{0.469} &\textbf{0.585} & 0.861\\
					&&&& \\
					n-mixed &A2C& 0.669 &0.382& \textbf{0.885}& 0.452 \\
					\bottomrule
				\end{tabular}
			\end{sc}
		\end{small}
	\end{center}
	\vskip -0.1in
\end{table}

\subsection{Alternate reward function on DS2}
\label{sec:rewards}

In this section we explore the performance of A2C when using a more sophisticated reward function (Equation \ref{eq:reward}). These experiments used DS2 and were benchmarked against an LSTM. Results are given in Table \ref{tab:newreward}. The A2C with access to only a single $s_t$ is able to outperform the LSTM with window size 20 by a notable margin. When the LSTM is given access to an even larger window size it gains the advantage on most performance metrics, but still fails to match the sensitivity of the A2C algorithm. This suggests a possibly interesting follow-up study where window data is provided to the A2C algorithm as well. 

Overall, the more complex reward function based on vitals improved the sensitivity versus the simpler function used in Tables \ref{tbl:compareagents} and \ref{tbl:bm1}. This suggests the possibility that a better reward function could be found in future work. 

\begin{table}[t]
	\caption{Comparing the performance results of A2C agents with the new reward function using DS2}
	\label{tab:newreward}
	\vskip 0.13in
	\begin{center}
		\begin{small}
			\begin{sc}
				\begin{tabular}{lccccc}
					\toprule
					Window & Algo. & AUC & MCC & Sens. & Spec. \\
					\midrule
					&A2C& 0.619 & 0.309 & \textbf{0.928}& 0.310 \\
					w-20&LSTM& 0.597 &0.237& 0.894& 0.300 \\
					w-30&LSTM& \textbf{0.700}&\textbf{0.395} & 0.825& \textbf{0.575}\\
					\bottomrule
				\end{tabular}
			\end{sc}
		\end{small}
	\end{center}
	\vskip -0.1in
\end{table}

\section{Conclusion}
\label{sec:conclusion}

In this work we design and demonstrate the application of deep reinforcement learning to the healthcare domain. Supervised and semi-supervised approaches to false alarm detection require feature engineering and domain expertise to scale and generalize, which is data intensive and expensive. Our RL approach mimics medical domain experts with high sensitivity, while still catching a notable number of false alarms. To the best of our knowledge this is the first application of deep RL for medical alarm annotations.

Agents trained with our mixed downsampling technique have superior performance compared with any single downsampling approach in this highly imbalanced data set. Our optimal A2C configuration achieved 88.5\% sensitivity in detecting true alarms, and 45.2\% specificity in identifying false alarms. This was accomplished after training on only one week's worth of scanner data. 

Identifying clinically false alarms, as done here, is a significantly more difficult task than identifying false alarms caused by artifacts and technical errors. Clinical reasoning requires an understanding of the patients' high-level physiological state as well as significant domain knowledge. Expert annotators used 20 second windows of data to visualize and adjudicate the alarms. Our techniques were competitive with LSTMs which simulated that view of the data, while requiring only a single data point. We believe that expanding RL techniques to the broader data window could result in significantly improved performance. It may also be interesting to represent the signals visually before feeding them into a hybrid CNN/RL algorithm. 

We would like to extend this work to specific alarm types (emergent, urgent, indeterminate) and refine the reward function in our future work. Our current work, however, can be applied to any event detection problem that follows a Markov decision process.

The limitations of this work are: i) training and testing of the RL agent is limited to one week of data; ii) we could find no prior work employing deep RL techniques on false alarm detection to benchmark against; iii) the alarm detection task is limited to two classes. 

The main contributions of this work are: i) A2C performs and generalizes better than DQN; ii) Adam optimizer is more stable than RMSProp for this problem; iii) mixed sampling ranges perform well for RL state representation compared to uniform downsampling; iv) our approach is data efficient, scalable to multiple tasks, and not very compute intensive. Furthermore, such methods could soon enable many practical non-clinical applications such as easing the data annotation processes thereby allowing faster generation of new healthcare datasets.



\end{document}